\documentclass[preprint, pra, amsmath, aps,showpacs]{revtex4}
\usepackage[dvips]{color}
\usepackage[usenames]{xcolor}
\usepackage{ulem}

\newcommand{\bea}{\begin{eqnarray} }
\newcommand{\eea}{\end{eqnarray}}
\newcommand{\bean}{\begin{eqnarray*}}
\newcommand{\eean}{\end{eqnarray*}}
\newcommand{\nn}{\nonumber \\}
\def\od#1,#2{\frac{d#1}{d#2}}
\def\odz#1,#2{\frac{d^2#1}{d{#2}^2}}
\def\pd#1,#2{\frac{\partial #1}{\partial #2}}
\def\pdz#1,#2{\frac{\partial^2 #1}{\partial {#2}^2}}
\def\pdd#1,#2{\frac{\partial^3 #1}{\partial {#2}^3}}
\def\pdv#1,#2{\frac{\partial^4 #1}{\partial {#2}^4}}
\def\pdzz#1,#2,#3{\frac{\partial^2 #1}{\partial {#2}\partial{#3}}}

\def\eq#1{Eq.~(\ref{#1})}

\def\eqn#1{(\ref{#1})}

\def \ni{\noindent}

\def\A{{\bf A}}

\def\B{{\bf B}}

\def\c{{\bf c}}

\def\J{{\bf J}}

\def\jac{{\cal J}}

\def\s{{\bf s}}
\def\S{{\bf S}}
\def\u{{\bf u}}

\def\btimes{~{\bf \times}~}
\def\bnabla{{\bf \nabla}}
\def\bcdot{~{\bf \cdot}~}
\def\ni{\noindent}

\newcommand{\lbs}{\left (}
\newcommand{\rbs}{\right )}

\newcommand{\lbm}{\left\lbrack}
\newcommand{\rbm}{\right\rbrack}
\newcommand{\lbl}{\left \{ }
\newcommand{\rbl}{\right \} }

\def\od#1,#2{\frac{d#1}{d#2}}
\def\hod#1,#2{\frac{\hat d#1}{d#2}}
\def\odz#1,#2{\frac{d^2#1}{d{#2}^2}}
\def\odd#1,#2{\frac{d^3#1}{d{#2}^3}}
\def\pd#1,#2{\frac{\partial #1}{\partial #2}}
\def\pdz#1,#2{\frac{\partial^2 #1}{\partial {#2}^2}}
\def\pdd#1,#2{\frac{\partial^3 #1}{\partial {#2}^3}}
\def\pdv#1,#2{\frac{\partial^4 #1}{\partial {#2}^4}}
\def\pdzz#1,#2,#3{\frac{\partial^2 #1}{\partial {#2}\partial{#3}}}

\usepackage{graphicx}
\usepackage{dcolumn}
\usepackage{bm}
\usepackage{epsfig}

\begin{document}


%

\bibliographystyle{unsrt}
%
%
\title{ATEQ: Adaptive Toroidal Equilibrium code}
%
%
%
\author{Linjin Zheng,$^{a)}$\footnote{Corresponding author, email: lzheng@austin.utexas.edu} M. T. Kotschenreuther,$^{a)}$ F. L. Waelbroeck,$^{a)}$
and Y. Todo$^{b)}$} 

\affiliation{$^{a)}$Institute for Fusion Studies, University of Texas at Austin, Austin, TX 78712\\
$^{b)}$National Institute for Fusion Science, National Institutes of Natural Sciences, Toki, Gifu 509-5292, Japan}

\date{\today}

\begin{abstract}
A radially adaptive numerical scheme is developed to solve  the Grad-Shafranov equation for axisymmetric magnetohydrodynamic equilibrium.  A decomposition with independent solutions is employed in the radial direction and Fourier decomposition is used in the poloidal direction. The independent solutions are then obtained using an adaptive shooting scheme together with the multi-region matching technique in  the radial direction. 
Accordingly,  the Adaptive Toroidal Equilibrium  (ATEQ) code is constructed for axisymmetric equilibrium studies. 
The adaptive numerical scheme in the radial direction improves considerably the accuracy of the equilibrium solution.  The decomposition with independent solutions effectively reduces the matrix size in solving the magnetohydrodynamic equilibrium problem.
 The reduction of the matrix size is about an order of magnitude as compared with the conventional radially grid-based numerical schemes. Also,  in this ATEQ numerical scheme, no matter how accuracy in the radial direction is imposed, the size of the matrices 
basically does not change.  The small matrix size scheme gives ATEQ more flexibility to address the requirement of the number of Fourier components in the poloidal direction in the tough equilibrium problems.  These two unique
 features, the adaptive shooting and small matrix size,  make ATEQ useful to improve tokamak equilibrium solutions.
 
 \vspace*{3mm}
\ni Key words: MHD, Equilibrium, Grad-Shafranov equation, tokamak, mirror

\end{abstract}
\pacs{52.35.Py, 52.55.Fa, 52.55.Hc}

\maketitle

\section{Introduction}

Solving the magnetohydrodynamic (MHD) equilibrium problem is fundamental in plasma physics and plays an essential role, in particular, in the magnetic confinement approach to fusion. In axisymmetric geometries, the equilibrium problem is reduced to solving the Grad-Shafranov equation \cite{grad,shaf}. Since this equation is nonlinear, a numerical solution is necessary in general. In advanced tokamaks \cite{bttry21}, two circumstances conspire to make the solution of the Grad-Shafranov equation particularly challenging. First, advanced tokamaks rely on broad current distributions to increase $\beta$, the ratio of kinetic to magnetic pressure.  This leads to a current profile peak near the edge and to the sensitivity of the stability limit to details in the geometry of the plasma edge. Second, they rely on the H-mode for confinement. The pressure gradients in H-mode drive localized, peaked bootstrap currents near the edge that add to the difficulty in two ways, first by increasing the stiffness of the Grad-Shafranov problem and second by increasing the accuracy needed to calculate the stability of Edge Localized Modes (ELM) \cite{ELITE2} as well as Resistive Wall Modes (RWM) \cite{ZKvD-RWM}.

Great efforts have been made previously to develop numerical solvers for the Grad-Shafranov equation. The applications for these numerical solvers are diverse, ranging from the interpretation of experimental observations \cite{efit,liuque} to the design of operation scenarios \cite{fable}, real-time control of experiments \cite{liuque,gpugss,pefit}, the analysis of the stability and transport properties of various configurations \cite{predict1}, and the optimization of machine designs \cite{maxj,bttry21}. The diversity of the applications leads to different requirements regarding properties of the algorithm such as speed, accuracy, stability, and flexibility. These different requirements are partly responsible for the multiplicity of solution strategies. The 1991 review article by Takeda and Tokuda  \cite{take} describes early codes including J-Solver \cite{jsol}, VMEC \cite{vmec}, TOQ \cite{toq}, and others  \cite{a1}-\cite{a4}.
Subsequent efforts led to the development of the codes CHEASE \cite{chea}, CORSICA \cite{cors}, and EFIT \cite{efit}.
Refs. \cite{a5}-\cite{a11} describe further works. As reviewed in  \cite{take}, the methods for solving the Grad-Shafranov equation are categorized into two types: the Eulerian or ``direct,'' and the Lagrangian or ``inverse'' numerical schemes.  The finite difference,  finite element, and  Fourier decomposition methods are employed to discretize the equation. In all cases, iteration 
is used to handle the nonlinearity.

Despite the great successes achieved with the existing codes in various scenarios, challenges remain for
solving  the equilibrium problem, especially  for the cases with high beta, strong shaping, and diverter geometries that give rise to separatrices. The need for adaptive solvers was realized a long time ago. It has, for example,  led to the development of the 
VMEC code for 3D equilibria \cite{vmec}. Later, the edge equilibrium code (EEC) was developed in order  
to address the numerical challenges pertaining to the tokamak edge equilibrium problem
\cite{eec}.

In this work, we introduce a new adaptive numerical scheme to solve the Grad-Schafranov equation
and describe its implementation in the ATEQ (Adaptive Toroidal EQuilibrium) code for tokamaks. The code uses a decomposition with independent solutions in the radial direction and Fourier decomposition in the poloidal direction. It then obtains the independent solutions with adaptive shooting  together with the multi-region matching technique in  the radial direction. 
The adaptive numerical scheme in the radial direction improves considerably the accuracy of the equilibrium solution.  The decomposition with independent solutions effectively reduces the matrix size in solving the magnetohydrodynamic equilibrium problem. 
The adaptive numerical scheme  has been successfully used in the linear 
MHD and kinetic stability codes, AEGIS \cite{aegis} and AEGIS-K \cite{aegisk}.

 In addition to its adaptive nature, the reduction
of the matrix size by ATEQ is about an order of magnitude, 
as compared to the conventional radially grid-based numerical schemes.
Also, in this ATEQ numerical scheme, no matter how accuracy
 in the radial direction is imposed, the size of the matrices basically does not change. 
Note that all numerical schemes for solving the Grad-Shafranov equation ultimately reduce to solving matrix equations. The size of matrices then matters. To achieve high accuracy, especially for tough problems related to the axis, X-point, or pedestal, etc. one has to increase the grid density in the radial and poloidal directions in the grid-based codes, or the radial grid density and the number of poloidal Fourier components in the Fourier-decomposition based codes. 
The dramatic reduction of matrix size by ATEQ is important for this research.
The Fourier-decomposition based codes remain to be important tools in this field,
for example  CORSICA is used for ITER, VMEC is still  popular.
The small matrix size scheme gives ATEQ more flexibility to address the requirement of the number of Fourier components
 in the poloidal direction for tough equilibrium problems.

The remainder of this paper is organized as follows: Sec.~II introduces the MHD equilibrium equations; 
Sec.~III describes the formulation of numerical equations; Sec. IV gives the numerical procedure and results; 
 Sec. V presents the benchmark studies and comparison with the existing equilibrium codes; Lastly, Sec. VI presents the conclusions and discussion.

\section{MHD Equilibrium equations}

In this section, we describe the MHD equilibrium equations and the goal of this work. Force balance, Amp\'{e}re's law, and the absence of magnetic charge form the basic set of equations describing
the MHD equilibrium for a static plasma (${\bf V}=0$) \cite{frei}:
\bea
&&\J\btimes\B=\bnabla p,
\label{eq1}
\\
&&\bnabla\btimes\B=\mu_0\J,
\label{eq2}
\\
&&\bnabla\bcdot\B=0,
\label{eq3}
\eea
where $\B$ is the magnetic field, $\J$ represents the current density, $p$ denotes the pressure,  $\mu_0$ is
the magnetic constant, and boldface denotes the vectors. 

The paper addresses axisymmetric toroidal equilibria. 
For such equilibria it is convenient to use a cylindrical  coordinate system $(X,Z,\phi)$, where $\phi$ is the toroidal angle, $Z$ denotes vertical coordinate,
 and $X$ is radial coordinate from the toroidal axisymmetric axis on the $\phi=0$ plane.  In this coordinate system, the magnetic field
in the axisymmetric case can be represented as \cite{frei}:
\bea
\B&=&\bnabla\phi\btimes\bnabla\chi+g\bnabla\phi,
\label{magb}
\eea
where $\chi$ is the poloidal magnetic flux. Both pressure $p(\chi)$ and $g(\chi)$ are flux functions.   

 Using the representation in \eq{magb} and equilibrium equations \eqn{eq1}-\eqn{eq3}, one can derive
 the so-called   Grad-Shafranov equation \cite{grad,shaf}. 
\bea
X^2\bnabla\bcdot\frac{\bnabla\chi}{X^2}&=& -\mu_0 X^2p'
-gg',
\label{gs}
\eea
where prime denotes the derivative with respect to the poloidal flux $\chi$. The MHD equilibrium is fully determined by $\chi$. 

The two free functions $p(\chi)$ and $g(\chi)$ need to be specified to determine $\chi$ from \eq{gs}. 
In practice, one usually specify $p$ and $g$ as the functions of  normalized
flux  $\hat\chi = \chi/\chi_a$, where $\chi_a $ is the poloidal flux at the edge or on the last closed 
flux surface and the poloidal flux is assumed to be zero at the magnetic axis.

The goal of the present paper is to lay out a new numerical scheme to solve \eq{gs} and describe the ATEQ code that implements this scheme.
The paper restricts attention to the fixed boundary problem, i.e., solving \eq{gs} with the plasma boundary specified. We defer consideration of the free boundary problem to future work.
 
\section{Formulation of numerical equations}
 
 \label{seciii}
 
In this section, we describe the numerical scheme to solve the Grad-Shafranov equation \eqn{gs} with the fixed boundary condition. 
We begin by describing the decomposition of the Grad-Shafranov equation before giving the computation of the metric parameters. We then describe the iteration scheme and boundary conditions.
We conclude this section with the description of the numerical scheme to solve  the equilibrium equations with the independent solution decomposition in the radial direction and the Fourier decomposition in the poloidal direction.
 
\subsection{Decomposition of the Grad-Shafranov equation}
\label{ssecdec}
 
 In this subsection, we introduce the radial, poloidal, and toroidal coordinates and project the Grad-Shafranov equation 
 onto this  coordinate system. We then use Fourier decomposition to decompose the equations.

 To solve the Grad-Shafranov equation, we introduce the coordinate system 
 $(\psi,\theta,\phi)$, with $\psi$ labelling the radial  grids  and $\theta$ being
 the poloidal angle. The coordinates $\psi$ and $\theta$ is general, only requiring that
 the Jacobian 
 \bean
 \jac &=& \frac1{\bnabla\psi\btimes\bnabla\theta \bcdot\bnabla\phi}
 \eean
 remains finite. In this coordinate system one can obtain
\bean
\frac1{X^2}\bnabla ~\chi &=& \frac1{X^2}\pd  ~\chi,\psi \bnabla\psi +\frac1{X^2}\pd  ~\chi,\theta \bnabla\theta
\nn
&=& A_1 \bnabla\theta\btimes \bnabla \phi+ A_2\bnabla\phi\btimes \bnabla \psi,
\eean
where
\bea
A_1&=& \frac1{X^2}\pd ~ \chi,\psi \jac |\bnabla\psi|^2+\frac1{X^2}\pd ~\chi,\theta  \jac\bnabla\psi\bcdot\bnabla\theta,
\label{a1}
\\
A_2&=& \frac1{X^2}\pd ~\chi,\psi \jac \bnabla\psi\bcdot\bnabla\theta +\frac1{X^2}\pd ~\chi,\theta\jac |\bnabla\theta|^2.
\nonumber
\eea
Therefore, one  has
\bean
\jac\bnabla\bcdot \frac {\bnabla ~\chi}{X^2}&=&\pd A_1,\psi +\pd A_2,\theta 
\nn
&=&\pd A_1,\psi 
+i {\cal M}\lbs \frac1{X^2} \jac\bnabla\psi\bcdot\bnabla\theta\rbs \pd ~\chi,\psi
- {\cal M}\lbl \frac1{X^2}\jac |\bnabla\theta|^2\rbl {\cal M}\chi.
\eean
Here, we have denoted $\pd,\theta = i{\cal M}$ with ${\cal M}$ being the matrix specifying the poloidal Fourier numbers, since the Fourier decomposition with $\theta$ will be
introduced later on. Using this decomposition the Grad-Schafranov equation \eqn{gs}
can be reduced to the set of first order differential equations:
\bea
\pd \chi,\psi&=& F_{11}\chi + F_{12}A_1,
\label{eq10}
\\
\pd  A_1,\psi &=&F_{21}\chi + F_{22}A_1+ S,
\label{eq20}
\eea
where
\bean
F_{11}(\psi, \theta)  &=& - i\frac{\bnabla\psi\bcdot\bnabla\theta}{|\bnabla\psi|^2} {\cal M},
\\
F_{12}(\psi, \theta)  &=& \frac{X^2}{\jac |\bnabla\psi|^2} ,
\\
F_{21}(\psi, \theta) &=&  {\cal M} \frac1{X^2}\jac |\bnabla\theta|^2 {\cal M}  - {\cal M}\frac1{X^2} \jac\bnabla\psi\bcdot\bnabla\theta\frac{\bnabla\psi\bcdot\bnabla\theta}{|\bnabla\psi|^2} {\cal M}
\\
F_{22}(\psi, \theta) &=& -i{\cal M}\frac1{X^2} \jac\bnabla\psi\bcdot\bnabla\theta \frac{X^2}{\jac |\bnabla\psi|^2} ,
\\
S(\psi, \theta) &=& \frac {1}{\chi_a}\lbs - \jac ~ {p}_{\hat\chi}'
-\frac\jac{ X^2} g {g}'_{\hat\chi}\rbs. 
\eean

To solve the set of equilibrium equations, Eqs. \eqn{eq10} and \eqn{eq20}, the following Fourier 
decompositions are introduced,
\bea
\begin{pmatrix}
 \chi \\ A_1\\S
\end{pmatrix}
&=& \frac1{\sqrt{2\pi}} \sum_{m=-M_{max}}^{M_{max}} 
\begin{pmatrix}
 \chi_m \\ A_{1m}\\ S_m
\end{pmatrix} \exp\{i  m\theta\}.
\label{four}
\eea
Here, $M_{max}$ represents the maximum Fourier component to be used. Introducing the Fourier
decomposition in \eq{four} 
the set of equations \eqn{eq10} and \eqn{eq20} becomes the set of matrix equations with the coefficients
becoming the maxtrices as defined as follows
\bean
 {\cal F}_{ij,mm'}
&=& \frac1{2\pi}\int_{-\pi}^\pi d\theta
F_{ij}(\psi,\theta)  e^{i (m'-m)\theta}.
\eean
Note that for the non-up-down symmetric system the Fourier components are complex. The set of
 matrix equations in complex can be written as
\bea
\pd,\psi \begin{pmatrix}
\chi \\ A_1
\end{pmatrix}
-
\begin{pmatrix}
 {\cal F}_{11}& {\cal F}_{12}
\\
 {\cal F}_{21}& {\cal F}_{22}
\end{pmatrix}
\begin{pmatrix}
\chi
 \\A_1 
\end{pmatrix}
&=&
\begin{pmatrix}
0
\\
 S 
\end{pmatrix}
\label{eqmat}.
\eea
Here, $\chi$ and $A_1$ are the vectors in the Fourier space with the total components $M = 2m_{max}+1$ 
for each and 
$ {\cal F}_{ij}$ are the matrices with dimension $M\times M$.  Therefore,
\eq{eqmat} represents a set of $2M$  differential equations. The matrix equation, \eq{eqmat}, can
be rewritten concisely as follows
\bea
\pd\u,\psi - {\cal F} {\u}=\s(\u),
\label{equ0}
\eea
where the source term $\s$ is usually a nonlinear function of $\u$.
  
\subsection{Computation of the metric parameters}
\label{sseccom}

In this  subsection, we describe how the matrix ${\cal F}$ is computed in the ATEQ code. 
This is related to the determination of
 the metric parameters, such as $|\bnabla \psi|^2$, $\bnabla \psi\bcdot\bnabla\theta$, etc. 
 
As in the PEST code \cite{pest},  we introduce the polar coordinates to compute the metric parameters:
\bean
S_r&=&x^2+z^2,
\nn
\Theta&=& \arctan(z/x),
\eean
where $x=X-X_0$ and $z=Z$ with $X_0$ being the major radius at the magnetic axis locating at $Z=0$. Noting that, since $X(\psi,\theta)$ and $Z(\psi,\theta)$ are given when introducing the ($\psi,\theta$) grids,  one can also determine $S_r(\psi,\theta)$
and $\Theta(\psi,\theta)$. Consequently, one can derive both
\bean
\frac{\partial(S_r,\Theta)}{\partial(X,Z)}&\equiv&
\begin{pmatrix}
\pd S_r,X &\pd S_r,Z
\\
\pd \Theta,X&\pd \Theta,Z
\end{pmatrix} ~~~\text{and}~~~
\frac{\partial(S_r,\Theta)}{\partial(\psi,\theta)}\equiv
\begin{pmatrix}
\pd S_r,\psi &\pd S_r,\theta
\\
\pd \Theta,\psi &\pd \Theta,\theta
\end{pmatrix}.
\eean
Using these results one can compute the metric parameters in the $(S_r,\Theta,\phi)$ coordinate system.

We first work on the
 Jacobian $\jac$. Note that
\bean
\jac &=&\frac{1}{\bnabla\psi\btimes\bnabla\theta\bcdot\bnabla\phi}
\nn
&=&\frac{X}{ \left|\frac{\partial(\psi, \theta)}{\partial(X,Z)}\right|}
=X \left|\frac{\partial(X,Z)}{\partial(\psi, \theta)}\right|.
\eean
Note further that
\bean
\frac{\partial(S_r,\Theta)}{\partial(\psi,\theta)}&=&
\frac{\partial(S_r,\Theta)}{\partial(X,Z)}
\frac{\partial(X,Z)}{\partial(\psi,\theta)}
~~~\text{and}~~~
\left| \frac{\partial(S_r,\Theta)}{\partial(X,Z)}\right|=2.
\eean
One  obtains the Jacobian expression in the polar coodinates 
\bean
\jac&=&X \left|\frac{\partial(X,Z)}{\partial(\psi,\theta)}\right|
=\frac X 2
 \left|\frac{\partial(S_r,\Theta)}{\partial(\psi,\theta)}\right|.
\eean

Next, we work on other metric parameters. By straightforward reduction one can obtain
\bean
\pd {(\psi,\theta)},{(X,Z)} &=&  \pd {(\psi,\theta)},{(S,\Theta)} \pd {(S,\Theta)},{(X,Z)}
\nn
&=& \lbs \pd {(S,\Theta)},{ (\psi,\theta)}\rbs^{-1} \pd {(S,\Theta)},{(X,Z)}
\nn
&=&
\frac{X}{ \jac}
\begin{pmatrix}
z \pd S_r,\theta/2S_r + x \pd \Theta, \theta& 
-x\pd S_r,\theta/2S_r + z \pd \Theta, \theta
\\
-z \pd S_r,\psi/2S_r -x \pd \Theta,\psi &
x \pd S_r,\psi/2S_r- z \pd \Theta,\psi
\end{pmatrix},
\eean
where it has been noted that
\bean
\begin{pmatrix}
\pd S_r,\psi    & \pd S_r,\theta  \\
\pd \Theta,\psi & \pd \Theta, \theta
\end{pmatrix}^{-1}
&=&\frac X{2 \jac} 
\begin{pmatrix}
 \pd \Theta, \theta    & -\pd S_r,\theta  \\
-\pd \Theta,\psi &\pd S_r,\psi
\end{pmatrix}.
\eean
Noting further that
\bean
|\bnabla\psi|^2&=&\pd\psi,X \pd\psi,X+ \pd \psi,Z\pd \psi,Z,
\\
|\bnabla\theta|^2&=&\pd\theta,X \pd\theta,X+ \pd \theta,Z\pd \theta,Z,
\\
\bnabla\psi\bcdot\bnabla\theta&=&\pd\psi,X \pd\theta,X+ \pd \psi,Z\pd \theta,Z,
\eean
one obtains
\bean
|\bnabla\psi|^2
&=&
\lbs\frac{X}{\jac}\rbs^2\lbm \frac1{4S_r} \lbs \pd S_r,{\theta}\rbs^2
+S_r \lbs \pd \Theta, {\theta}\rbs^2\rbm,
\\
|\bnabla\theta|^2&=&\lbs\frac{X}{\jac}\rbs^2\lbm \frac1{4S_r} \lbs \pd S_r,{\psi}\rbs^2
+S_r \lbs \pd \Theta, {\psi}\rbs^2\rbm,
\\
\bnabla\psi\bcdot \bnabla\theta
&=&\lbs\frac{X}{\jac}\rbs^2\lbm
-\frac1{4S_r} \pd S_r,\psi\pd S_r,{\theta} 
-S_r \pd \Theta,\psi\pd \Theta, {\theta}\rbm.
\eean
Using the toroidal symmetry property, we can also find that
\bean
|\bnabla\phi|^2=\frac1{X^2},
~~~
\bnabla\psi\bcdot\bnabla\phi= 0,
~~~\text{and}~~~
\bnabla\theta\bcdot\bnabla\phi
&=&0.
\eean

The expressions of metric parameters given above can be used to compute the matrix ${\cal F}$ and the vector
$\s$ in \eq{equ0}.

\subsection{Iteration scheme and boundary conditions}

\label{ssecite}

With the computation of metric parameters given in the last subsection, we describe the iteration scheme to solve 
the Grad-Shafranov equation with proper boundary conditions.

Since the equation, \eq{equ0}, are nonlinear, an iteration process is necessary. One can follow the usual
iteration scheme to get the converged solution:
 \bea
\pd \u^{(n+1)},\psi - {\cal F} \u^{(n+1)}=\s(\u^{(n)}).
\label{equ}
\eea
Here, $n$ denotes the iteration step.

Equation \eqn{equ} is a set of inhomogeneous differential equations of first order.  Its general solutions 
at step $n+1$ can be expressed as 
\bea
\u&=& \sum_{k=1}^{2M} c_k \u^k +\u^s,
\eea
where $c^k$ are the complex constants to be determined  by the boundary conditions,
$\u^k$ are the independent solutions to the homogeneous equations and $\u^s$  is the specific solution to
take into account the source term $\s$ on the right hand side of \eq{equ}. For brevity, the step index $n$ has been dropped.

Since the number of equations is $2M$, the solutions are completely determined by the $M$
boundary conditions in complex at the magnetic axis and $M$ boundary conditions in complex at plasma edge $\chi_a$. 
The boundary conditions at plasma edge $\chi_a$ are specified by the given shape of the
last closed flux surface in the fixed boundary value problem. The boundary conditions 
at the magnetic axis are just the requirement that the independent solutions are  ``small" in terms of
the  terminology of differential equation theory.  
The ``large" solution causes the system energy to diverge, while the ``small" solution is square-integrable with respect to the energy integral.  Near the magnetic axis, the homogeneous part of the Grad-Shafranov equation can be approximated by
the cylinder model. In this limit the solutions are as follows \cite{frei}:
\bean
\chi_m= &a_m r^{|m|} + b_m r^{-|m|}, ~~~ &for~m\not=0,
\\
\chi_0= &a_0+b_0\ln r,~~~& for ~m=0,
\eean 
where $r$ is the minor radius and $a_m$ and $b_m$ are constants. Therefore, the boundary conditions
for small solutions are simply $b_m=0$. This yields
\bea
\frac {d \chi_m/dr}{\chi_m}= & |m| r^{|m|-1}, ~~~ &for~m\not=0,
\label{bd1}
\\
\frac {d \chi_0/dr}{\chi_0}=&0,~~~& for ~m=0.
\label{bd2}
\eea
 The boundary conditions for $A_{1m}$ can be obtained using the definition of $A_1$ in \eq{a1}.
 
  Note that the general solution to the set of differential equations is the summation
 of homogeneous solutions and specific solution and
 the boundary conditions are satisfied by the constants $c_k$ tied to the 
 homogeneous solutions. Therefore,  the boundary conditions for specific solution are arbitrary.

\subsection{Solution of equilibrium equations}
\label{ssecsol}

The principle to solve \eq{equ} is laid out in  subsection \ref{ssecite}.  The actual implementation
is more complicated. One needs to divide multiple regions in the radial direction and then match the solutions 
in the individual regions to get the global solution. In this subsection, we will outline 
the actual numerical process
in the ATEQ code to solve the Grad-Shafranov equation.

 The $M$ boundary conditions at the magnetic axis in Eqs. \eqn{bd1} and \eqn{bd2} can be used to eliminate $M$ independent
solutions by shooting  outwardly with the boundary conditions at the axis
as the initial conditions. There are  only $M$ independent solutions $\u^k$ left as a result. In principle, the remaining  $M$ constants
$c_k$ can be determined by the other $M$ boundary conditions at the plasma edge, while the specific solution
$\u^s$ can be also determined by the numerical shooting with the boundary condition at the magnetic
axis $\u^s(0) =0$. The procedure  looks straightforward. However, this straightforward procedure to shoot all the way 
from the axis to the edge usually does not work due to the numerical pollution of large solutions. One has to divide the whole region into
multiple regions and then match the solutions in the individual regions to get the global solutions. 
A similar numerical scheme has been successfully used in the MHD stability code  AEGIS code \cite{aegis}.
 
Suppose there  are $L$ regions with their boundaries labeled respectively as $\psi_l$ $(l=0,1,2,\cdots,L)$,
where $\psi_0$ and $\psi_L$ represent respectively the magnetic axis and the last closed flux surface. 
For the first region, one can shoot from $\psi_0$ with the boundary conditions at the magnetic axis to get
$M$ independent solutions. The $M$ independent solutions at the other end $\psi_1$ are used the construct 
the independent solution  matrix:  $^1{\cal U}^{2M\times M} = (\u^1,\cdots,\u^M)_{\psi=\psi_1}$. 
Here, the left superscript indicates the region and the right superscript
 $2M\times M$ represents ``the number of Fourier components" $\times$ ``the number of independent solutions".
For the last region, one can shoot inwardly with the boundary conditions at $\psi_L$, i.e., by specifying $\chi$ at the edge,  to get
$M$ independent solutions. Likewise, the $M$ independent solutions can be used
to form the independent solution matrix at the other end $\psi_{L-1}$: $^L{\cal U}^{M\times2M}_{inward}$. Here,
the subscript ``inward" has been introduced to indicate the shooting in the last region is made inwardly. 
For the internal regions ($l=2,\cdots,L-1$) there are $2M$ independent solutions in each region. To construct
the independent solutions in these regions, for example, region $l$,  one can obtain
the $2M$ independent solutions by specifying the independent boundary conditions 
at the lower end $\psi_{l-1}$ and shooting upwardly.  The $2M$  independent boundary conditions 
at the lower end can be simply the $2M$ columns in the  identity matrix: ${\cal 
I}^{2M\times2M}$. The $2M$ independent solutions at the upper end of each internal region can
be used to form the independent solution matrices: $^l{\cal U}^{2M\times2M}$. 
 
Similarly, one can construct the specific solution vectors. In difference from the homogeneous solutions, 
there is only one set of solutions with $2M$ elements $^l\s^{2M\times 1}$ in each region. They can be obtained by
specifying the initial conditions at the lower end as the null vector  $0^{2M\times 1}$ (vector with all elements being zero) and shooting to the upper end, except the last region, The inward shooting is carried out in the last region.

With the independent solution matrices and the specific solution vectors in each region obtained, 
one can match them to obtain the global solutions. There are $L-1$ regional interfaces and on each 
interface there are $2M$ matching conditions. Note that, since the boundary conditions at axis and plasma edge have 
been applied,  only $M$ constants in each of these two regions remain to be determined. They
are represented in vectors of $M$ rows: $^1\c^{M\times1}$ and $^L\c^{M\times1}$.
In the internal regions, however, there are $2M$ constants in each region, which
are denoted as $^l\c^{2M\times 1}$ $(l=2,\cdots, L-1)$. 
Therefore, The $2M(L-1)$ matching conditions determine fully the constants $^l\c^{M_l}$,
 where $l=1,2,\cdots, L$ and $M_1=M_L=M$, $M_l=2M$ for $l\not =1$ and $L$.
The matching conditions can be expressed as follows
\bea
{\cal Y}
\begin{pmatrix}
^1\c^{M\times 1}
\\
^2\c^{2M\times 1}
\\
\vdots
\\
^{L-1}\c^{2M\times 1}
\\
^{L}\c^{M\times 1}
\end{pmatrix}^{2M(L-1) \times 1}
=
\begin{pmatrix}
-^1\s^{2M\times 1}
\\
-^2\s^{2M\times 1}
\\
\vdots
\\
-{^{L-2}\s}^{2M\times 1}
\\
{^L\s}^{2M\times 1}_{inward}-{^{L-1}\s}^{2M\times 1}
\end{pmatrix}^{2M(L-1) \times 1},
\label{cin}
\eea
where
\bean
{\cal Y}
\equiv
\begin{pmatrix}
^1{\cal U}^{2M\times M}&-{\cal 
I}^{2M\times2M}&0^{2M\times2M}&\cdots&0^{2M\times M}&0^{2M\times M}&0^{2M\times M}
\\
0^{2M\times M}&^2{\cal U}^{2M\times2M}&-{\cal 
I}^{2M\times2M}&\cdots&0^{2M\times M}&0^{2M\times M}&0^{2M\times M}
\\
\vdots&\vdots&\vdots&\vdots&\vdots&\vdots&\vdots
\\
0^{2M\times M}&\cdots&0^{2M\times2M}&\cdots&^{L-2}{\cal 
U}^{2M\times2M}&-^{L-1}{\cal I}^{2M\times 2M} &0^{2M\times M}
\\
0^{2M\times M}&\cdots&0^{2M\times2M}&\cdots&0^{2M\times M}&^{L-1}{\cal 
U}^{2M\times2M}&-^L{\cal U}^{2M\times M}_{inward}
\end{pmatrix}^{2M(L-1) \times 2M(L-1)}
\eean
Matrix ${\cal Y}$ is a band matrix. By inverting it one can obtain  the solution of \eq{cin} 
\bea
\begin{pmatrix}
^1\c^{M\times 1}
\\
^2\c^{2M\times 1}
\\
\vdots
\\
^{L-1}\c^{2M\times 1}
\\
^{L}\c^{M\times 1}
\end{pmatrix}^{2M(L-1) \times 1}
= {\cal Y}^{-1}
\begin{pmatrix}
-^1\s^{2M\times 1}
\\
-^2\s^{2M\times 1}
\\
\vdots
\\
-{^{L-2}\s}^{2M\times 1}
\\
{^L\s}^{2M\times 1}_{inward}-{^{L-1}\s}^{2M\times 1}
\end{pmatrix}^{2M(L-1) \times 1},
\label{scin}
\eea
With the constants obtained from \eq{scin}, the solutions in each region are then simply
\bea
^l\u&=& \sum_{k=1}^{M_l} ~^lc_k ~^l\u^k +~^l\u^s,~~~(l=1,\cdots,L).
\label{global}
\eea
These give the numerical scheme being implemented in the ATEQ code to solve the Grad-Shafranov equation.

\section{Numerical procedure and results}

In this section, we describe how to implement  the numerical scheme in section \ref{seciii}. 
This leads to the development of  ATEQ code. The computational flow chart of ATEQ is given in Fig.~1. 
To be more specific to describe the computational flow, we use an ITER-like equilibrium as an example.
The major radius 6.2 m, minor radius 2 m, elongation 1.78, triangularity 0.4, 
 the vacuum magnetic field at the geometric center of plasma column is $6$ T,  the total current $15.9$ MA,  
and the volume average beta value is
$3.371\%$. Figure 2 shows the cross section with the ``a" part showing the initial grid setup and the ``b" part 
showing the magnetic surfaces computed by the ATEQ code. The case will also be used for the benchmark studies 
with the TOQ code. Further details about the equilibrium will be described there.

 First, one needs to set up radial and poloidal grids 
 ($\psi,\theta$) as shown in Fig. 2a. The grids are constructed to surround the magnetic axis  ($x_{axis}, z_{axis}$). Because the
 magnetic axis is unknown beforehand, iteration is needed. 
 The value of the previous step ($n$) 
 is used to construct the grids to advance to the next step $n+1$.  Following the iteration scheme in \eq{equ}, 
the source term on the right hand side of  \eq{equ} is evaluated by using the solution for poloidal flux 
$\u^{(n)}(\psi^{(n)},\theta^{(n)})$ in the previous step. Note that the pressure and current profiles are prescribed
by the normalized poloidal flux. The total poloidal flux $\chi_a$ needs also to be determined iteratively. 
At the first step, the quantities at the previous step are prescribed
by initial guessing. 
The matrices ${\cal F}$ and $\s$ can then be computed with previous step grids as described 
in subsection \ref{sseccom}. Using the splines the matrices ${\cal F}$ and $\s$ are made to be radially continuous functions.

 Here, it is noted that the proper choice of initial ($\psi,\theta$) grids can affect how many Fourier components are
 required. For the usual equilibria without X points included the choice is rather arbitrary, i.e., a wide range of grid choices can work
 well.  For the equilibria with X points included proper choice of initial grids is important. In the ATEQ code, 
 the initial grids are specified as follows. First, the grids with ellipticity $k$ and triangularity $\delta$ are set up inside the specified
 plasma-vacuum boundary according to the formula
 \bea
X&=& x_{axis} + r(\cos\theta -\delta\sin^2\theta),
\label{initgx}
\\
Z&=& z_{axis} + k r\sin\theta
\label{initgz}
\eea
with $r=[(X-x_{axis} )^2+(Z-z_{axis} )^2]^{1/2}$.  Here, $k$ and $\delta$ can be polynomial  functions of $\psi$.  This means that
one can adjust the ellipticity and triangularity from the axis to the outmost surface. In  most cases,  the linear dependence 
is sufficient. Next, the difference between the specified plasma boundary and the outmost surface given by Eqs. \eqn{initgx}
and \eqn{initgz} are distributed radially. The distribution can be adjusted through an exponential multiplier of $\psi$. 
Also, the $\psi$ grids can be packed near the axis and boundary.
In our experience, with these flexibilities, roughly 100 Fourier sidebands are sufficient to get a good equilibrium 
solution with X points included. It is using this type of initial grid setting that the Solov\'ev solution with X points to be described later is reproduced numerically. There is always a possibility to use the ($\psi,\theta$) solution at step $n$ for the grids
at  step $n+1$. Nevertheless, it can only be used if the solution at step $n$  is sufficiently smooth and well-behaved.

Next, the whole radial domain is split into $L$ regions.  As described in subsection \ref{ssecsol}, adaptive
shooting is implemented to get the independent solution matrices in each region. By solving 
for $^l\c^{M_l}$  using \eq{scin}, one can construct the global solution through \eq{global}. At this step, we first check if
the  magnetic axis ($x_{axis}, z_{axis}$)   and total poloidal flux $\chi_a$ converge. Usually,  total poloidal flux converges in one or two steps, using the following formula  for prediction
\bean
\chi_a^{(n+2)} = \sqrt{ \chi_a^{(n+1)}\chi_a^{(n)}}.
\eean
Instead, to find  the  magnetic axis  ($x_{axis}, z_{axis}$) one needs a few iterations.  
 The code  shoots outwardly  from the  assumed  magnetic axis ($x_{axis}, z_{axis}$). After achieving 
the solution, the minimum of poloidal flux $\chi$ is determined. The location of this minimum is used as 
($x_{axis}, z_{axis}$) for the shooting in the next step. This process is repeated until the starting ($x_{axis}, z_{axis}$) matches
the location of the $\chi$ minimum  computed to a required accuracy.

Figure 3 shows the iteration process for determining
 the magnetic axis to get the final solution in Fig. 2b. 
 The dashed curve in Fig. 3a shows the poloidal magnetic flux on the mid-plane computed with 
  a guess value of magnetic axis
 in an earlier step.  By searching for the minimum of the poloidal flux, a new magnetic axis location 
 is found as shown by the solid vertical line.
It is iterated until the magnetic axis coordinates ($x_{axis},z_{axis}$) converge. Figure 3b shows the  converged result.
 Because the iteration for magnetic
axis and the iteration for the overall solution of poloidal magnetic flux are implemented simultaneously, the overall
solution is often converged as the axis searching converges. The requirement for the number 
of poloidal Fourier components 
is also verified. The example shown in Fig. 2 uses 50 sidebands. 

With the  magnetic axis ($x_{axis}, z_{axis}$)   and total poloidal flux $\chi_a$ being converged, one can further iterate 
 to get the converged solution $\chi(\psi,\theta)$. With this solution, one can obtain the numerical solution
 for the poloidal magnetic flux $\chi(X, Z)$. The magnetic surfaces with $\chi(X, Z)=const$ are plotted in Fig. 2b.

\section{Benchmark studies and comparison with existing codes }
\label{benchmark}

This section describes the benchmark studies. We begin with the analytical Solov\'ev equilibrium
 with the X points included \cite{solo}. Next, the comparison with the existing numerical
equilibrium code TOQ \cite{toq} is detailed.
 We also build a backward substitution module to double check the numerical equilibrium solutions. It simply substitutes
the solution $\chi(X,Z)$ back into the Grad-Shafranov equation to check if the equation is satisfied to a sufficient accuracy.   
The convergency and comparisons with other codes are vindicated by the check with the backward substitution 

module to double check the numerical equilibrium solutions. It simply substitutes
the solution $\chi(X,Z)$ back into the Grad-Shafranov equation to check if the equation is satisfied to a sufficient accuracy.

The benchmark with the Solov\'ev solution is not a trivial task. This is because the X-points are present in the Solov\'ev equilibrium. 
The equilibrium computation with the X-points included is challenging because much more
Fourier components  are needed.  The analytical Solov\'ev solution is given as follows
\bea
\chi_{Solovev} &=&\frac12 \lbs bX_0^2+c_0X^2\rbs Z^2 +\frac18(a-c_0)(X^2-X_0^2)^2,
\label{Solovev}
\eea
where the parameters are given as follows in the benchmark studies: $X_0=10$, $a=1$, $b=-0.83$, and $c_0=0.92$. This solution corresponds to the pressure and poloidal
current flux profiles given as follows
\bea
-p' =a~~~ \hbox{and}~~~-gg'/X_0^2 = b.
\label{solpf}
\eea
As pointed out in Ref.  \cite{solo}, the second-order solution in \eq{Solovev}
is actually an exact solution of the Grad-Shafranov equation. 

The numerical procedure for the benchmark studies to the analytical Solov\'ev equilibrium solution is as follows. From  the Solov\'ev solution in \eq{Solovev} one can determine the last closed flux surface.
The last closed flux surface is then used as the plasma boundary condition in the ATEQ code. The same pressure and
poloidal current flux profiles as given in \eq{solpf} are used to compute the    numerical solution $\chi(X,Z)$ with the ATEQ code. The  solution is then  compared with the analytical Solov\'ev solution in \eq{Solovev}.

Figure 4a gives the initial  ($\psi,\theta$)   grids and Fig. 4b shows the converged magnetic flux surfaces
computed by the ATEQ code.
The number of Fourier sidebands is $102$.
The process just follows the chart given in Fig. 1. The numerical results agree well with the
analytical solution in \eq{Solovev}.  To show the agreement, 
the poloidal magnetic flux at the mid-plane on the low field side both from the analytical solution in \eq{Solovev} (solid curve)
and from the computational result by ATEQ (dashed curve) are plotted in Fig. 5. Two curves completely overlap, although the initial guessing as shown in Fig. 4a deviates dramatically from the actual solution in Fig. 4b in the ATEQ computation.

Comparisons with the existing equilibrium codes are also performed. Here, we describe a benchmark example
 with the TOQ  equilibrium code \cite{toq}. A typical case is described 
as follows. A TOQ  sample initiation file with $equiltype='ffprime'$ is taken. 
To be more specific to compare with TOQ,
here the same numerical parameter notations as in the TOQ  manual are used to describe  the 
equilibrium parameters. 
The shape of boundary type is specified by $ishape=2$, 
which is described as follows:
\bea
X&=& rzero + rmax*(\cos\theta -xshape*\sin^2\theta),
\nn
Z&=& eshape* rmax*\sin\theta,
\eea
where the basic parameters are specified as follows: The major radius $rzero = 6.2~m$, the minor radius $rmax=2~m$, the elongagtion $eshape =1.78$, and the triangularity
$xshape=0.4$. This leads the equilibrium cross section to be given in Fig. 2.

The pressure gradient ($p'$) and poloidal current flux parameter ($gg'$) profiles are specified, respectively,
by setting $modelp=3$ and $modelf = 1$, which are described as follows:
\bea
p' &=& 1-0.4\hat\chi+0.4\hat\chi^2 - \hat\chi^3,
\label{pprof}
\\
gg' &=& 1-\hat\chi.
\label{gprof}
\eea
Note here that the profiles are specified with the normalized poloidal magnetic flux $\hat\chi$, varying from $0$ to $1$ from the magnetic axis to plasma boundary.
The tolerance is set to be $toleq=10^{-5}$ in the TOQ iteration with successively increasing grid densities. 
Here, we have used the nonlinear pressure profile in \eq{pprof}, which is different from the TOQ sample initiation file,
in order to avoid the linear profile case considered in the Solov\'ev case.  The pressure and poloidal current flux profiles
are given in Fig.  6 with respect to the minor radius on the outer vertical mid-plane.  Although the $p'$ and $gg'$ are the
same as specified in Eqs. \eqn{pprof} and \eqn{gprof} for TOQ and ATEQ codes, the pressure $(p)$ and poloidal current flux ($g$) profiles can be 
slightly different since they are given in the minor radius, instead of the normalized poloidal flux. The slight difference of poloidal magnetic flux solution as discussed later
can cause  the difference.   The volume average beta  is $3.371 \%$,    the normalized beta is  $2.54$, and $l_i =  0.730$ in
this equilibrium.

Figure 2b shows the equilibrium magnetic flux surfaces by the ATEQ code. 
The slight difference between TOQ and ATEQ results is not perceivable in the
flux surface plot. Figure 7  is introduced to show the poloidal magnetic flux $\chi$ and the safety factor $q$ versus 
the minor radius, which are computed, respectively,
by the TOQ (red) and ATEQ (blue) codes. One can see that both $\chi$ and $q$ solutions agree rather well. 
The slight difference results from the different accuracies for TOQ and ATEQ codes as discussed later on in
the backward substitution check.
Note that the red (TOQ) and blue (ATEQ) curves in Figs. 6 and 7
terminate roughly at the same minor radius. This indicates that the Shafranov shifts computed by the two codes agree. 

 In passing, it is pointed out that  the region number $L$ is about $30-40$ to recover the Solov\'ev solution.
For the case without X points, the required number $L$ is less. 
It usually does not work if $L=1$, i.e., shooting all the way from the axis to the plasma edge. 
Some Fourier components become extremely larger, while some others are very small. This feature
makes the final matching matrix at the edge in poor condition.  The multiple region matching
solves the difficulty. Because the matrix size in ATEQ is determined by the number of regions, instead
of the radial grid points, and the number of regions is much less than the grid points, 
adding some more regions does not cause many difficulties.

To further check the computation results, we implement the backward substitution check both for TOQ and
ATEQ. In this checking procedure, the numerical solution $\chi$ is substituted back to the Grad-Shafronov
equation,  \eq{gs}, to compute the relative errors at each grid point.  
Because the $\chi$ is determined,
$p$ and $g$ become one dimensional. The solution $(\chi,\theta)$ are used as the coordinates for checking. The 5 point differential scheme is used to evaluate the derivatives. This check is done surface by surface.
The relative error for each grid point 
is  defined by the difference between  the left
and right values divided by the larger one between them. 
The surface-averaged relative errors
are plotted in Fig. 8 versus the normalized magnetic flux. Because of the adaptive numerical scheme, high accuracy
or low relative error is achieved by the ATEQ calculation. For TOQ computation, a very low tolerance $toleq=10^{-5}$ has actually been
imposed. The TOQ code does exit with the  converged results. 
The convergence criterion in the TOQ code is based on the comparison between two consecutive steps, instead of the backward substitution check as in the ATEQ code. This explains the larger surface-averaged relative error as compared to the ATEQ code in the backward substitution check.  We especially want to emphasize that
this does  not necessarily imply that TOQ is not good, but only shows that different convergence criteria can 
yield different solutions. If TOQ used the backward substitution method to determine the convergence, TOQ could 
possibly also  get good results. 
 Also, the ATEQ code is based on an adaptive numerical scheme. Better convergence
can be expected. The backward substitution check of ATEQ results further
verifies its numerical procedure. 

We also performed checks with other codes, for example, VMEC and EFIT. ATEQ achieves satisfactory results, generally giving better convergence in the backward substitution check. The backward substitution method thus confirms the validity of the ATEQ code.

\section{Conclusions and discussion}

We have presented a new, radially adaptive numerical scheme that solves the Grad-Shafranov equation for axisymmetric  MHD equilibrium. This numerical scheme represents the solution through a sum in terms of independent solutions in the radial direction and Fourier decomposition in the poloidal direction. It computes the independent solutions using an adaptive shooting scheme together with the multi-region matching technique in  the radial direction. The adaptive numerical scheme improves considerably the accuracy of the equilibrium solution. We named the implementation of this scheme the Adaptive Toroidal EQuilibrium code (ATEQ). 

The decomposition with independent solutions effectively reduces the matrix size for solving the magnetohydrodynamic equilibrium problem, as compared with numerical schemes based on a fixed radial grid. The adaptive numerical scheme
 is expected to be especially helpful to deal with  stiff equilibrium problems. Our results also indicate that the backward substitution method
can be necessary to obtain a reliable equilibrium solution.

Let us here further discuss the unique features of the ATEQ numerical scheme.
The numerical methods for solving the Grad-Shafranov equation ultimately reduce the problem to solve 
the matrix equations. The matrix size then matters and reducing the matrix size in discretizing the Grad-Shafranov equation is
important.  
In the grid-based numerical schemes both in the radial and poloidal directions (finite difference or finite element), the size of the matrix is $N_r \times N_t$. Here, $N_r$ and $N_t $ are respectively the numbers of grids in the radial and poloidal directions. In the numerical scheme based on the poloidal Fourier decomposition, the matrix size is $N_r\times N_f$. Here, $N_f$ is the number of the poloidal Fourier components. To achieve high accuracy, especially for tough problems related to the axis, X-point, or pedestal, etc. one has to increase the $N_r$ and $N_t$ (or $N_f$). Consequently, the size of the matrices becomes large and the matriices become
hard to deal with numerically. In the ATEQ numerical scheme, the radial direction is split into $L$ regions with each region addressed by the adaptive shooting of independent solutions. It reduces the radial $N_r$ grid problem into a $L$ region matching problem. This cuts down the $N_r\times N_t$ (or  $N_r\times N_f$) 
matrix problem in the conventional numerical schemes into a $L \times N_{indep}$ problem in the ATEQ numerical scheme. Here, the number of regions $L$ is about a few 10s and $N_{indep}$ is
the number of independent solutions, which is of the same order as $N_f$.  
The reduction of the matrix size is by the factor $L/N_r$, which is 
about an order of magnitude, as compared with the conventional radially grid-based numerical schemes. Also, in this ATEQ numerical scheme, no matter how accuracy in the radial direction is imposed, the size of the matrices basically does not change. 
Such an improvement in the order of magnitude rarely happens. 
It therefore represents a significant development in this research.

The equilibrium problem is a little bit different from the stability one. If one uses the exact flux solution as the radial grids, only a single Fourier component for the magnetic flux $\chi$ is required because it is constant on the surfaces. 
Therefore, the required number of the Fourier components, 
$N_f$, in principle can be somewhat optimized by setting proper radial grids. 
Since the matrix size is reduced in the radial direction in ATEQ, one  has more flexibility to increase 
the umber of poloidal Fourier components if it is required. This is a distinct feature of ATEQ as compared to the conventional Fourier decomposition based codes.  This improvement
is useful.

 It is realized in this field that a good numerical equilibrium solution near the axis and plasma boundary 
in the presence of the X points is critical. 
It is a challenging issue for decades. As cited in the introduction, several efforts have been made.  
Our work provides another possible solution. To directly compare with other codes to treat the X point equilibrium problem
will be our next task. This may require close collaboration with other teams. 
Equilibrium codes often need certain specific procedure  to execute them.
Using the backward substitution method we found that the equilibrium accuracy varies a lot even with the same code. 
That's why we're wary of doing code-to-code comparisons directly without the other party involved. Each code
may have their own particular features. We have compared with TOQ since the example file is in the public domain. Even in this
case, we have provided additional clarifications.  But, one thing we can do is to compare with the Solov\'ev analytical solution
in the presence of X points. If not at all, rather few codes have been published with such a comparison as justification.  
This shows the capacity of ATEQ numerical scheme and code.

This research is supported by the U. S. Department of Energy, Office of Fusion Energy Science
under Grant No. DE-FG02-04ER54742 and the  US-Japan Joint Institute for Fusion Theory (JIFT) collaboration program

\newpage

\ni Figure captions:

Fig. 1: Computational flow chart of the ATEQ code.

Fig. 2:  Equilibrium results for the ITER-geometry-like case. a) The initial ($\psi,\theta$)   grids; b) The converged magnetic flux surfaces.

Fig. 3: The iteration process to determine the magnetic axis. a) The initial guessing; b) The converged result.  
Dashed curves indicate the poloidal flux and the vertical solid lines indicate the proposed magnetic axis at the
respective iteration step. 

Fig. 4: Equilibrium results for Solov\'ev solution. a) The initial ($\psi,\theta$)   grids; b) The converged magnetic flux surfaces.

Fig. 5: The poloidal magnetic flux at mid-plane on the low field side both from the analytical Solov\'ev 
solution in \eq{Solovev} (solid curve)
and the ATEQ computational result (dashed curve). Two curves completely overlap.

Fig. 6: The equilibrium pressure and poloidal current flux profiles  versus the minor radius on the outer
 mid-plane for the benchmark case between TOQ (red) and ATEQ (blue).  

Fig. 7: The equilibrium poloidal magnetic flux $\chi$  and safety factor  profiles  versus the minor radius on the outer
 mid-plane for the benchmark case as computed by TOQ (red) and ATEQ (blue).  

Fig. 8:  The surface-averaged relative errors 
 versus the normalized magnetic flux with the backward substitution check, respectively, for TOQ (red) and ATEQ (blue) numerical results.

\newpage
\begin{figure}[htp]
\centering
\large
\includegraphics[width=80mm]{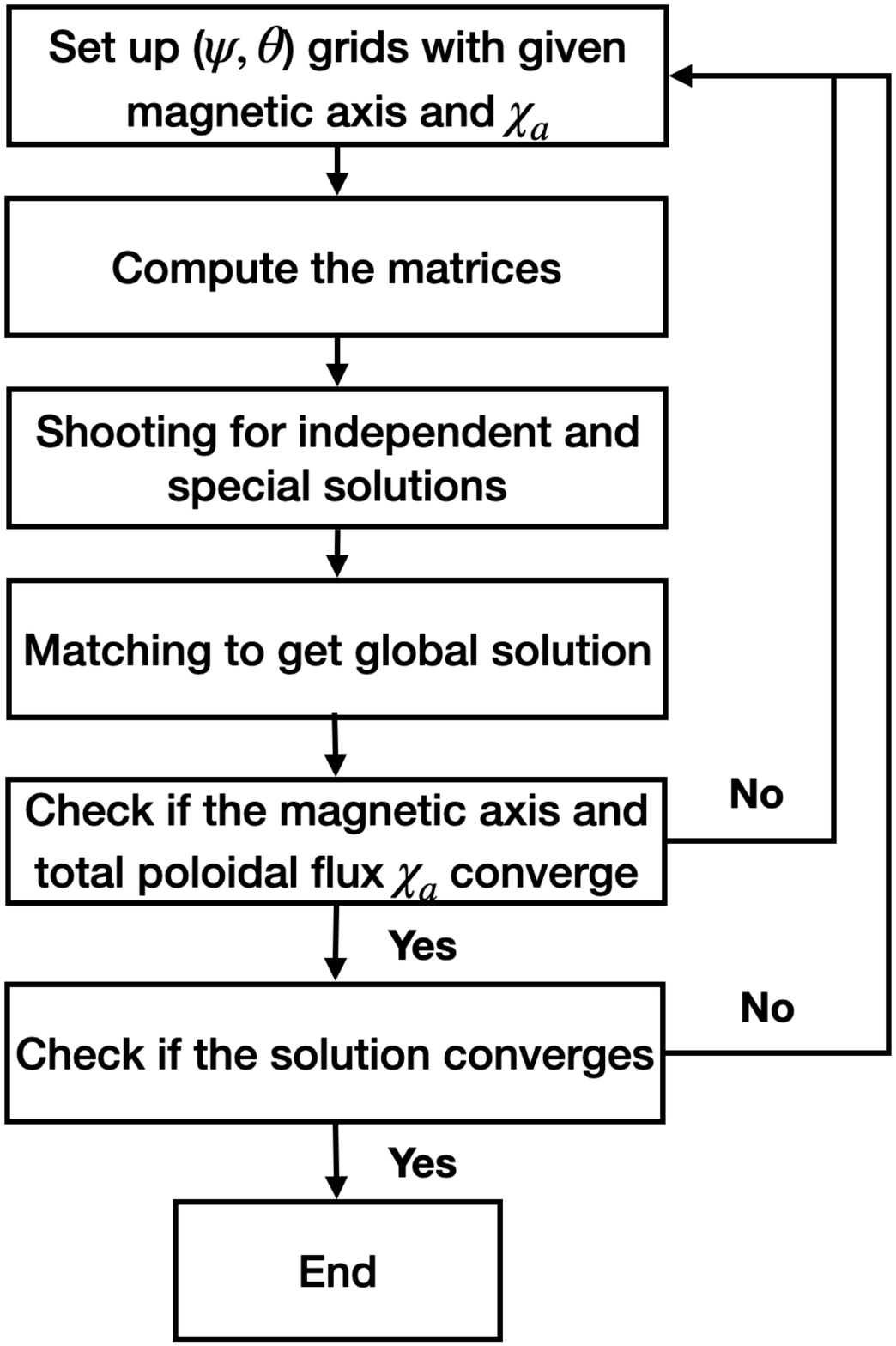}
\end{figure}

\vspace*{20mm}

\centerline{Fig. 1}

\newpage

\begin{figure}[htp]
\centering
\large
\includegraphics[width=130mm]{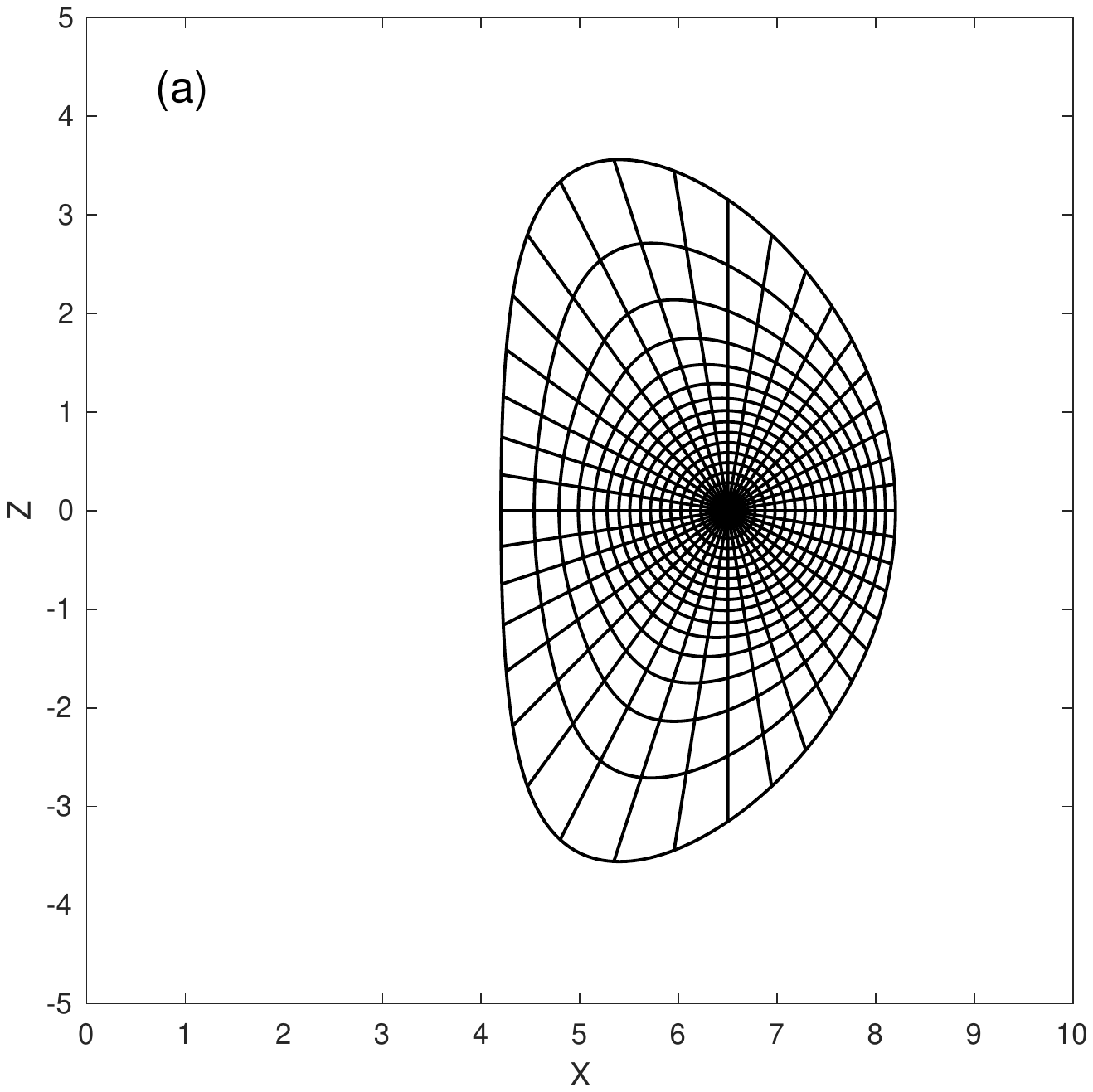}
\end{figure}

\vspace*{20mm}

\centerline{Fig. 2a}

\newpage
\begin{figure}[htp]
\centering
\large
\includegraphics[width=130mm]{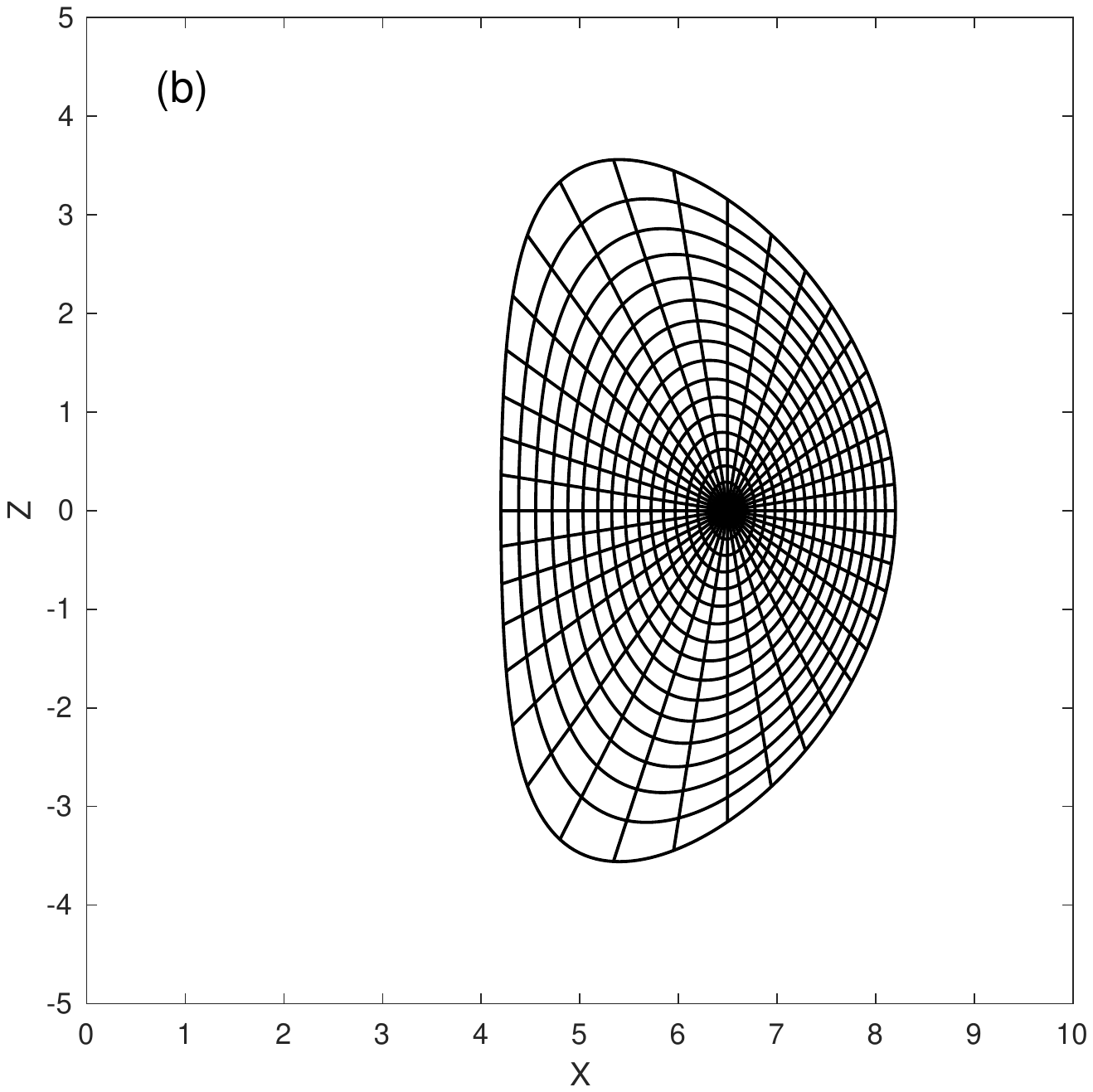}
\end{figure}

\vspace*{20mm}

\centerline{Fig. 2b}

\newpage
\begin{figure}[htp]
\centering
\large
\includegraphics[width=100mm,angle=-90]{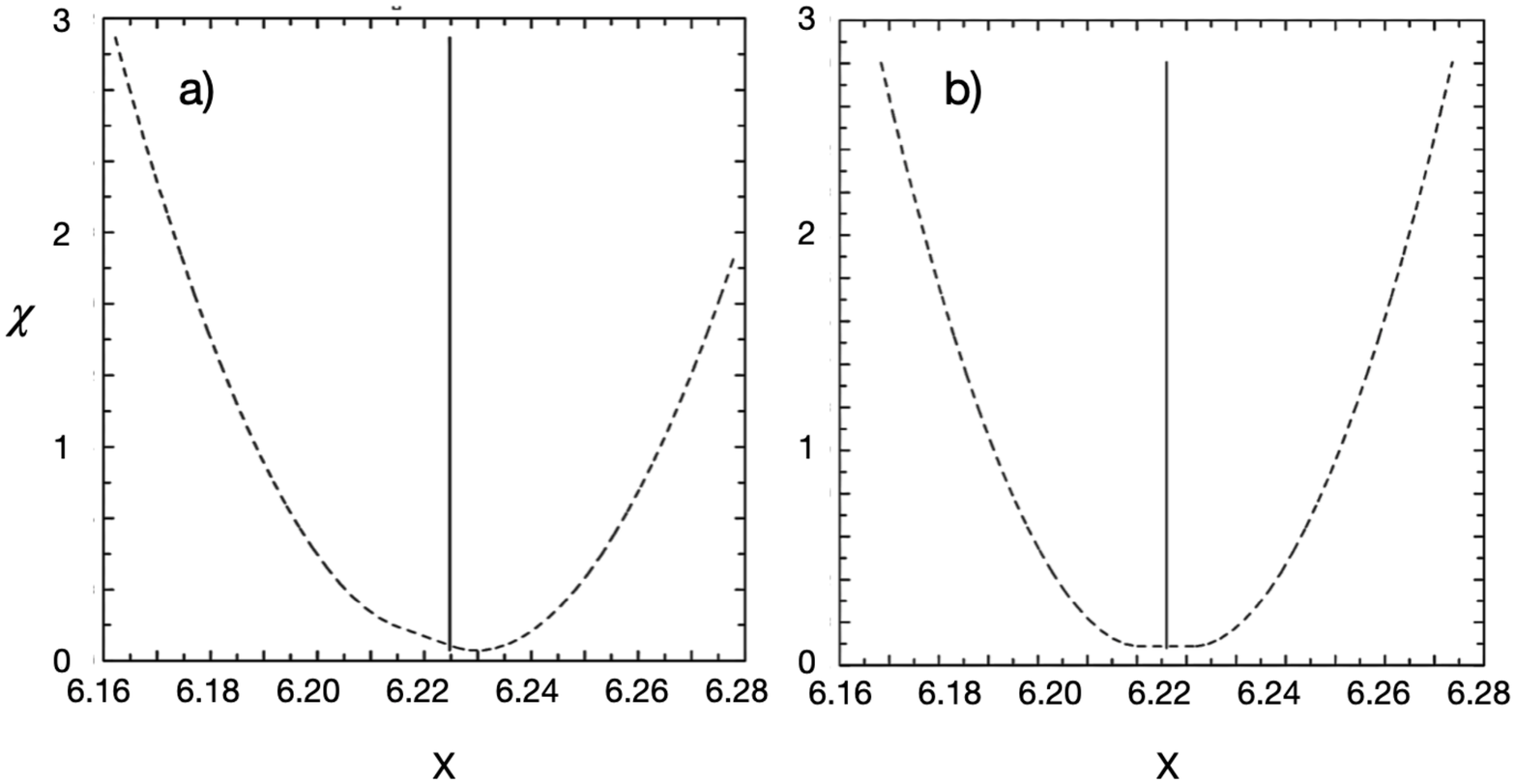}
\end{figure}

\vspace*{20mm}

\centerline{Fig. 3}

\newpage
\begin{figure}[htp]
\centering
\large
\includegraphics[width=100mm,angle=-90]{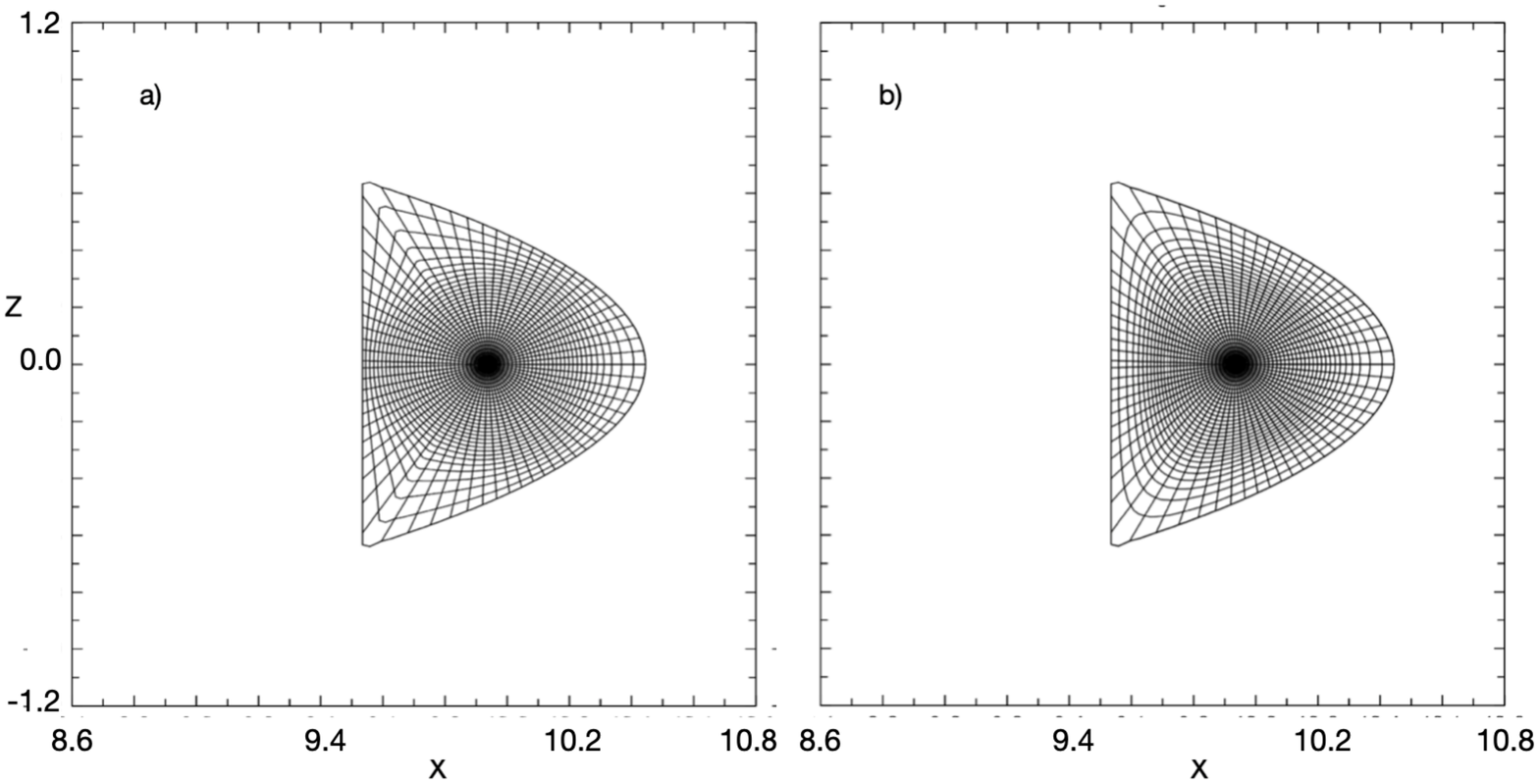}
\end{figure}

\vspace*{20mm}

\centerline{Fig. 4}

\newpage
\begin{figure}[htp]
\centering
\large
\includegraphics[width=90mm]{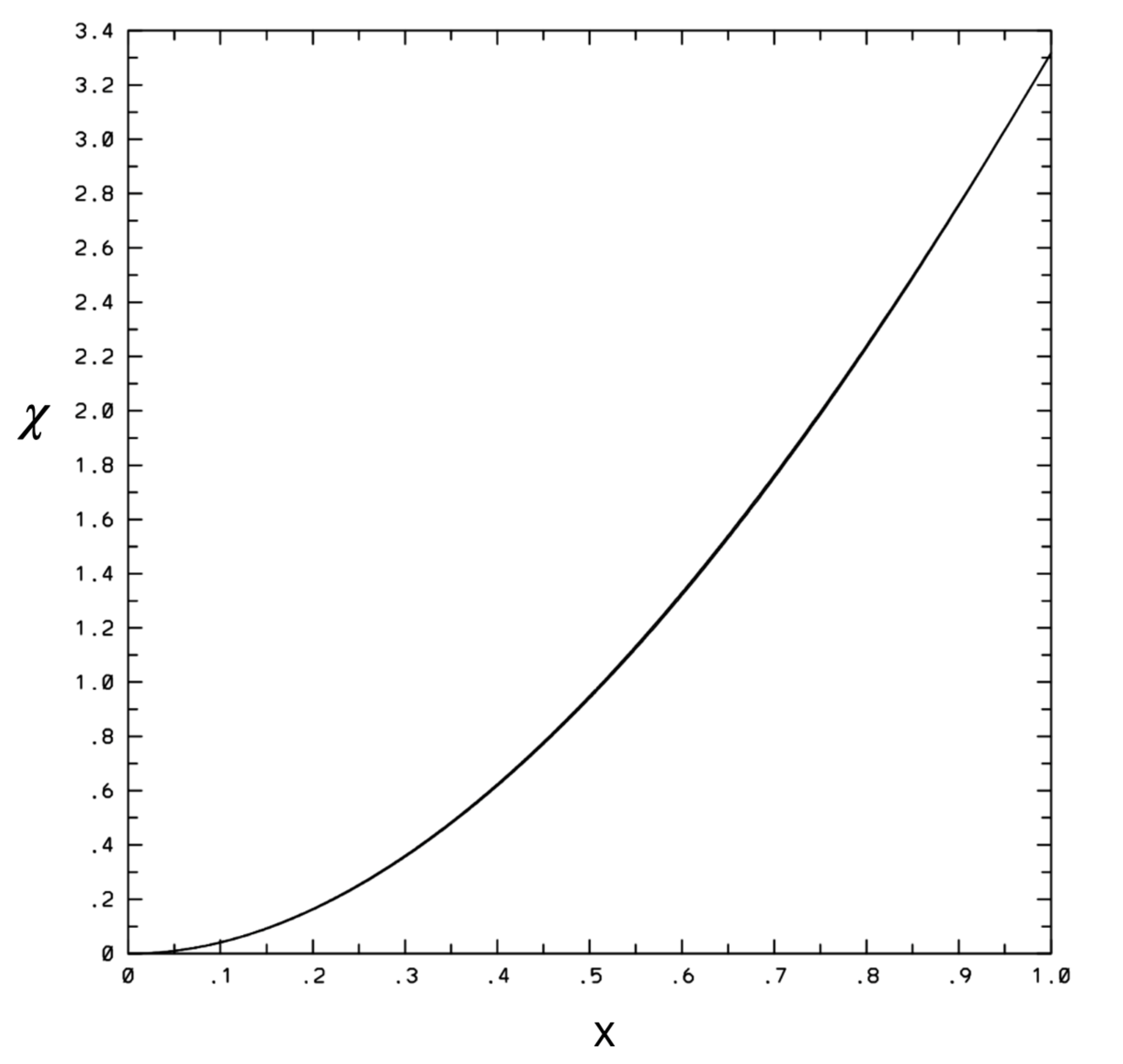}
\end{figure}

\vspace*{20mm}

\centerline{Fig. 5}

\newpage
\begin{figure}[htp]
\centering
\large
\includegraphics[width=120mm]{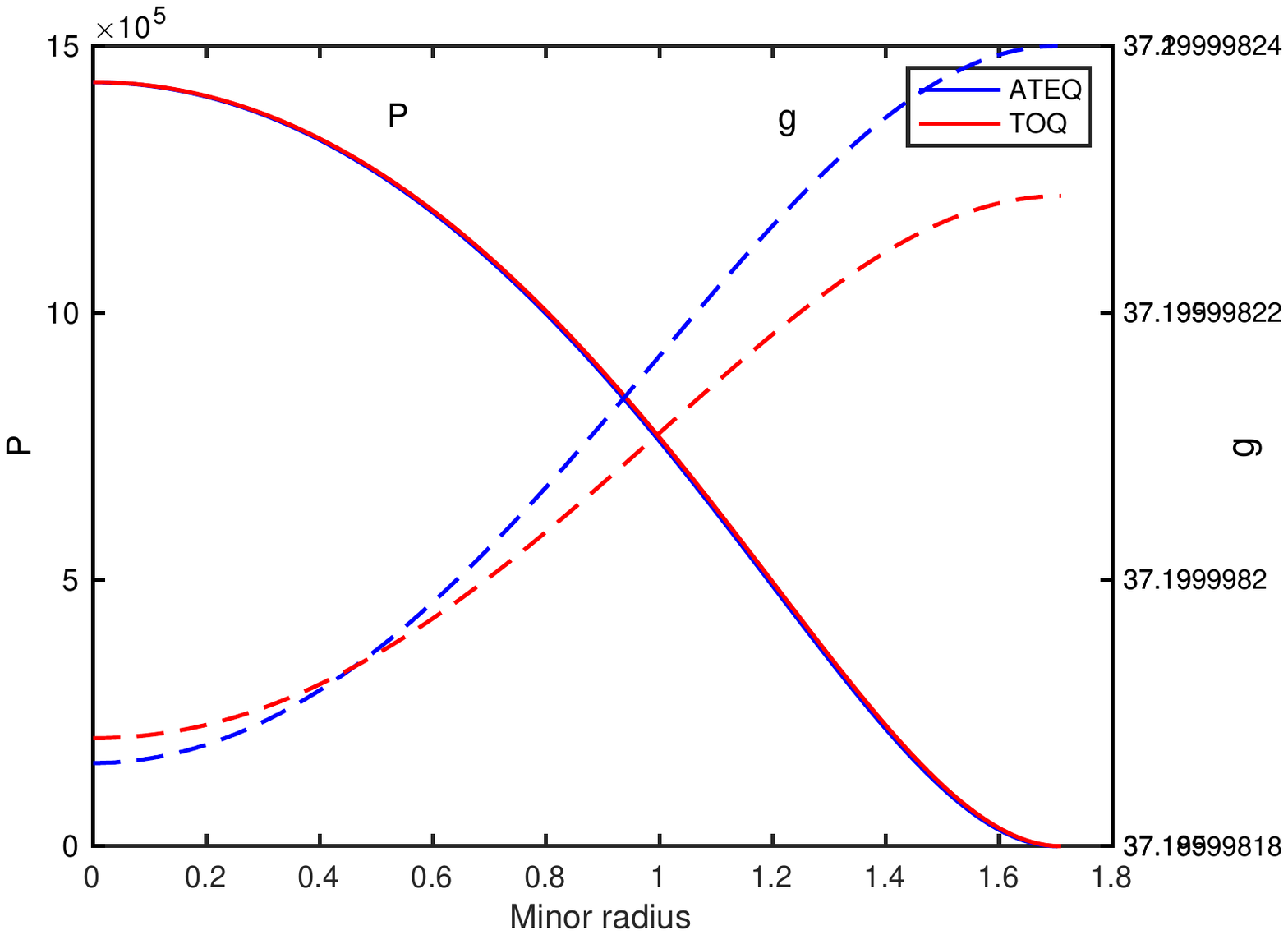}
\end{figure}

\vspace*{20mm}

\centerline{Fig. 6}

\newpage
\begin{figure}[htp]
\centering
\large
\includegraphics[width=100mm]{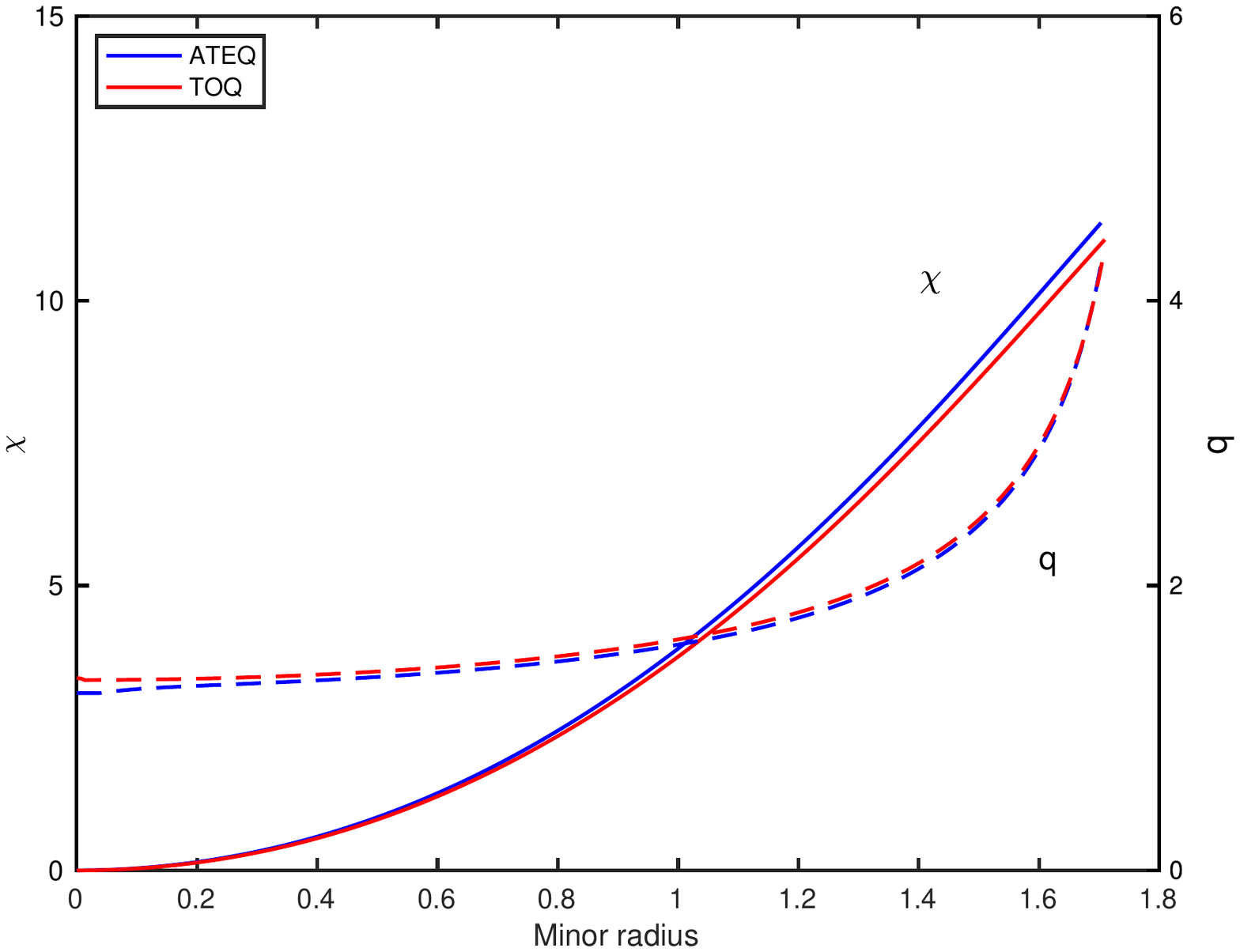}
\end{figure}

\vspace*{20mm}

\centerline{Fig. 7}

\newpage
\begin{figure}[htp]
\centering
\large
\includegraphics[width=110mm]{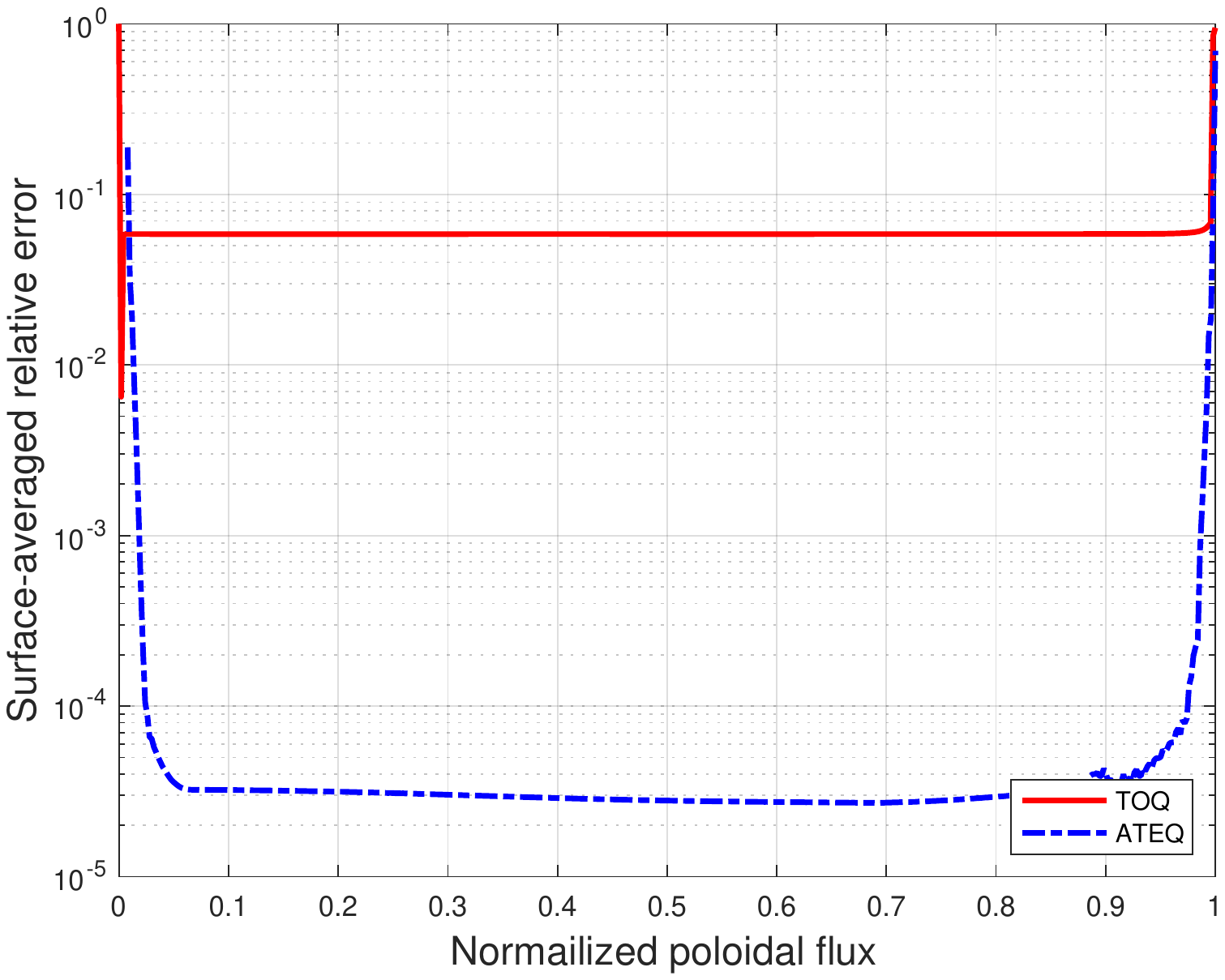}
\end{figure}

\vspace*{20mm}

\centerline{Fig. 8}

\newpage

\newpage


\begin{thebibliography}{10}

\bibitem{grad}
H. Grad and H. Rubin, Proc. 2nd Int. Conf. on the Peaceful Uses of Atomic Energy, Vol. 31 (United Nations, Geneva, 1958) p. 190-97. https://digitallibrary.un.org/record/3808553?ln=en.

\bibitem{shaf}
 V.D. Shafranov, ZhETF 33 (1957) 710-22; Sov. Phys. JETP 6 (1958) 545-54. http://www.jetp.ac.ru/cgi-bin/e/index/e/6/3/p545?a=list.

\bibitem{bttry21} 
R.J. Buttery, J.M. Park, J.T. McClenaghan, D. Weisberg, J. Canik, J. Ferron1, A. Garofalo, C.T. Holcomb, J. Leuer, P.B. Snyder, and The Atom Project Team, Nucl. Fusion  61 (2021) 046028. https://doi.org/10.1088/1741-4326/abe4af.

\bibitem{ELITE2} 
P.B. Snyder, K.H. Burrell, H.R. Wilson, M.S. Chu, M.E. Fenstermacher, A.W. Leonard, R.A. Moyer, T.H. Osborne, 
M. Umansky, W.P. West, and X.Q. Xu,
Nuclear Fusion 47 (2007) 961-968. https://doi.org/10.1088/0029-5515/47/8/030.

\bibitem{ZKvD-RWM}
L.J. Zheng, M.T. Kotschenreuther and J.W. Van Dam
Phys. Plasmas 17 (2010) 056104. https://doi.org/10.1063/1.3318267.


\bibitem{efit}
L.L. Lao, H. St. John, R.D. Stambaugh, A.G. Kellman and W. Pfeiffer,
``Reconstruction of current profile parameters and plasma shapes in tokamaks",
 Nucl. Fusion 25 (1985) 1611.  https://doi.org/10.1088/0029-5515/25/11/007.


\bibitem{liuque}
J.M. Moret, B.P. Duval, H.B. Le, S. Coda, F. Felici and H. Reimerdes, Fus. Eng. Design 91  (2015) 1 - 15.
https://doi.org/10.1016/j.fusengdes.2014.09.019.

\bibitem{fable}
E. Fable, C. Angioni, A. A. Ivanov, K. Lackner, O. Maj, S. Yu Medvedev, G. Pautasso, G. V. Pereverzev, W. Treutterer, and the ASDEX Upgrade Team, Plasma Phys. Control. Fusion 55 (2013) 074007. https://doi.org/10.1088/0741-3335/55/7/074007.

\bibitem{gpugss}
Y. Huang, B.-J. Xiao and Z.-P. Luo, Chinese Physics B 26  (2017) 085204-8. http://cpb.iphy.ac.cn/EN/10.1088/1674-1056/26/8/085204.

\bibitem{pefit}
X.N. Yue, B.J. Xiao, Z.P. Luo and Y. Guo, Plasma Phys. Control. Fusion 55 (2013) 085016. https://doi.org/10.1088/0741-3335/55/8/085016.

\bibitem{predict1}
 B. C. Lyons,  C. Paz-Soldan, O. Meneghini,  L. L. Lao, D. B. Weisberg,  E. A. Belli,  T. E. Evans,  N. M. Ferraro, and P. B. Snyder,
Phys. Plasmas 25 (2018) 056111. https://doi.org/10.1063/1.5025838.

\bibitem{maxj}
R. Miller, M. Chu, R. Dominguez and T. Ohkawa, Comments on Plasma Phys. Control. Fusion 12 (1989) 125-32.

\bibitem{take}
T. Takeda and S. Tokuda,
Journal of Computational Phys. 93 (1991) 1-107.
https://doi.org/10.1016/0021-9991(91)90074-U.

\bibitem{jsol}
J. DeLucia, S.C. Jardin, A.M.M. Todd, J. Comput. Phys. 37 (1980) 183-204.
https://doi.org/10.1016/0021-9991(80)90020-0

\bibitem{vmec}
S.P. Hirschman and J.C. Whitson, Phys. Fluids 26 (1983) 3553-68. https://doi.org/10.1063/1.864116

\bibitem{toq}
Originally written by Bob Miller of General Atomics, https://fusion.gat.com/THEORY/toq/overview.html.

\bibitem{a1}
J. DeLucia, S.C. Jardin, and A.M.M. Todd, J. Comput. Phys., 37 (1980) 183-204. https://doi.org/10.1016/0021-9991(80)90020-0.

\bibitem{a2}
J.A. Holmes, Y.K.M. Peng, and S. J. Lynch, J. Comput. Phys., 36 (1980) 35-54.
https://doi.org/10.1016/0021-9991(80)90173-4.

\bibitem{a3}
K.M. Ling and S.C. Jardin, J. Comput. Phys., 58 (1985) 300-35. 
https://doi.org/10.1016/0021-9991(85)90165-2.

\bibitem{a4}
R. Gruber, R. Iacono, and F. Troyon, J. Comput. Phys., 73 (1987) 168-82.
https://doi.org/10.1016/0021-9991(87)90112-4.

\bibitem{chea}
H. L\"{u}tjens, A. Bondeson, and O. Sauter,
Computer Physics Communications 97 (1996) 219-60.
https://doi.org/10.1016/0010-4655(96)00046-X.

\bibitem{cors}
J.A. Crotinger, L. LoDestro, L. Don Pearlstein, A. Tarditi, T.A. Casper, and E.B. Hooper,
 ``Corsica: A Comprehensive Magnetic-Fusion Devices
Simulation Final Report to the LDRD Program", 
UCRL-ID-1262S4 (1997). https://www.osti.gov/servlets/purl/522508.

\bibitem{a5}
G.T.A. Huysmans, J.P. Goedbloed, and W. Kerner, Proc. CP90 Conf. on Comp. Phys. Proc., pages 371-376, 1991.
https://doi.org/10.1142/S0129183191000512.

\bibitem{a6}
B. Turkington, A. Lifschitz, A. Eydeland, and J. Spruck, J. Comput. Phys., 106 (1993) 269-85.
https://doi.org/10.1016/S0021-9991(83)71107-1.

\bibitem{a6a}
L. Greengard and J.Y. Lee, J. Comp. Physics  125 (1996) 415-24.
https://doi.org/10.1006/jcph.1996.0103.

\bibitem{a7}
G.O. Ludwig, Plasma Phys. Controlled Fusion, 39 (1997) 2021-37.
https://doi.org/10.1088/0741-3335/39/12/006.

\bibitem{a8}
S.C. Jardin, J. Comput. Phys., 200 (2004) 133-52.
https://doi.org/10.1016/j.jcp.2004.04.004.

\bibitem{a8a}
L. Guazzotto, R. Betti, J. Manickam and S. Kaye,
Physics of Plasmas 11 (2004) 604-14. https://doi.org/10.1063/1.1637918.

\bibitem{a9}
P.A. Gourdain, J.N. Leboeuf, and R.Y. Neches,  J. Comput. Phys., 216 (2006) 275-99.
https://doi.org/10.1016/j.jcp.2005.12.005.

\bibitem{a10}
A. Pataki, A.J. Cerfon, J.P. Freidberg, L. Greengard,  and M. O'Neil, J. Comput. Phys. 243 (2013) 28-45.
https://doi.org/10.1016/j.jcp.2013.02.045.

\bibitem{a11b}
M.D. Boyer, D.J. Battaglia, D. Mueller, N. Eidietis, K. Erickson, J. Ferron, D.A. Gates, S. Gerhardt, R. Johnson, E. Kolemen, J. Menard,  et al., Nuclear Fusion  58 (2018)  036016.  
https://doi.org/10.1088/1741-4326/aaa4d0.

\bibitem{a11a}
T. S\'anchez-Vizuet and M. Solano, Comp. Phys. Communications  235  (2019) 120-32.
https://doi.org/10.1016/j.cpc.2018.09.013.

\bibitem{a11a1}
R. F. Schmitt, L. Guazzotto, H. Strauss, G.Y. Park, and C.S. Chang, 
Physics of Plasmas 18 (2011) 022502. https://doi.org/10.1063/1.3551714.

\bibitem{a11}
H. Li and P. Zhu, Computer Physics Communications, 260 (2021) 107264.
https://doi.org/10.1016/j.cpc.2020.107264.

\bibitem{eec}
X. Li, L.E. Zakharov, and V.V. Drozdov, Phys. Plasmas, 21 (2014) 012505.
https://doi.org/10.1063/1.4861369.

\bibitem{aegis}
L.J. Zheng, and M.T. Kotschenreuther,  
J. Comput. Phys. 211 (2006) 748-66.
https://doi.org/10.1016/j.jcp.2005.06.009.

\bibitem{aegisk}
L.J. Zheng, M.T. Kotschenreuther, and J.W.Van Dam, 
J. Comput. Phys. 229 (2010) 3605-22.
https://doi.org/10.1016/j.jcp.2010.01.017.

\bibitem{peel}
J.W. Connor, R.J. Hastie, and H.R. Wilson, Phys. Plasmas  5 (1998) 2687-700.
https://doi.org/10.1063/1.872956.

\bibitem{frei}
J.P. Freidberg, Ideal Magnetohydrodynamics (New York: Plenum Press, 1987).

\bibitem{pest}
R.C. Grimm, J.M. Greene, and J.L. Johnson,
   Methods of Computational Physics (Academic Press, New York,
   London, 1976). Vol.  9, p. 253-80.

\bibitem{solo}
L.S. Solov\'ev,
Sov. Phys. JETP 26 (1967) 400-7.
http://jetp.ac.ru/cgi-bin/e/index/e/26/2/p400?a=list.

\end{thebibliography}

\end{document}